\documentclass[aps,prd,preprint]{revtex4}
\usepackage{amsmath,amssymb}

\newcommand{\be}{\begin{equation}}
\newcommand{\ee}{\end{equation}}
\newcommand{\ea}{\end{array}}
\newcommand{\beqa}{\begin{eqnarray}}
\newcommand{\eeqa}{\end{eqnarray}}

\newcommand{\nn}{\nonumber}
\newcommand{\tr}{\mathop{\rm Tr}\nolimits}
\newcommand{\BI}{\openone}    
\newcommand{\half}{\frac{1}{2}}
\newcommand{\R}{{\rm I\!\rm R}}   
\newcommand{\C}{{\rm C}\llap{\vrule height6.3pt width1pt depth-.4pt\phantom t}}
\newcommand{\CP}{{\rm C}\llap{\vrule height6.3pt width1pt depth-.4pt\phantom t}{\rm P}}

\newcommand{\gapproxeq}{\lower .7ex\hbox{$\;\stackrel{\textstyle
>}{\sim}\;$}}
\newcommand{\lapproxeq}{\lower .7ex\hbox{$\;\stackrel{\textstyle
<}{\sim}\;$}}
\def\thebibliography#1{{\bf REFERENCES\markboth
 {REFERENCES}{REFERENCES}}\list
 {[\arabic{enumi}]}{\settowidth\labelwidth{[#1]}\leftmargin\labelwidth
 \advance\leftmargin\labelsep
 \usecounter{enumi}}
 \def\newblock{\hskip .11em plus .33em minus -.07em}
 \sloppy
 \sfcode`\.=10000\relax}

\begin{document}
\bibliographystyle{h-physrev}
\preprint{ SU-4252 797,\quad  DFUP-04-18  }
\title{Duality in Fuzzy Sigma Models   }
\author{A.P. Balachandran}
\affiliation{Physics Department, Syracuse University, 
Syracuse NY 13244-1130, USA}
 \email{bal@physics.syr.edu}
\author{ Giorgio Immirzi}
\affiliation{ Dipartimento di Fisica, Universit\`a di Perugia, \\
and INFN, Sezione di Perugia, Perugia, Italy}
 \email{giorgio.immirzi@pg.infn.it}

\begin{abstract}
Nonlinear `sigma' models in two dimensions have BPS solitons which are
solutions of self- and anti-self-duality constraints. In this paper, we
find their analogues for fuzzy sigma models on fuzzy spheres which were
treated in detail by us in earlier work. We show that fuzzy BPS
solitons are quantized versions of `Bott projectors', and construct
them explicitly. Their supersymmetric versions follow from the work of 
 S. K\" urk\c c\" uo\v glu.
\end{abstract}

\maketitle 

\section{Introduction}
\setcounter{equation}{0}

Nonlinear field theories such as $\CP^N$-models in two dimensions are of 
theoretical interest. For large $N$, $\CP^N$-models for example are 
asymptotially free and show features which resemble QCD. They can also
contain solitonic solutions and can thus serve as relatively simple
quantum field theories (qft's) for examining asymptotic freedom and solitons.

In Euclidean qft's in 2-d, spacetime can be compactified
to the two-sphere $S^2$. In turn $S^2$ can be discretized to the fuzzy sphere
$S^2_F$ by quantization \cite{madore} . Such a discretization, besides its novelty,
has other meritorious features such as preserving rotational invariance and
supersymmetry on $S^2$ \cite{kurki}\cite{kurkii}, avoiding fermion
doubling \cite{bali},
 and having a precise instanton, monopole and index theory \cite{balii}.
Numerical work on certain qft's on $S^2_F$ such as $(\phi^4)_2$ has also been
completed \cite{denjoe}. They do have correct limits to qft's on $S^2$ and
are thus alternatives to lattice regularization.

In previous work \cite{bagi}\cite{trg}, fuzzy nonlinear models, such as fuzzy
$\CP^N$
and Grassmanian models were constructed on $S^2_F$. Their supersymmetric
generalizations were also found \cite{kurkii}.

$\CP^1$ models on $S^2$ are of particular theoretical interest. As the target
space is $S^2$, they are models of ferromagnets. As shown by Belavin and
Polyakov \cite{belavin}, their solitons can be self-dual or anti-self-dual.
These solutions saturate a topological bound on actions and exactly solve the
field equations. They are the 2-d analogues of 4-d instantons.

In our work \cite{bagi}, we did not properly discuss fuzzy analogues of these
self- and anti-self-dual solutions. In this paper, we resolve this lack 
of completeness. We establish that the fuzzy $\sigma$-fields based on Bott
projectors are the fuzzy analogues of $S^2$-fields with a duality
invariance. As Kurk\c cuoglu's work \cite{kurkii} is  based on a 
supersymmetric generalization of Bott projectors, we now also have 
a supersymmetric version of these solitons.

\section{Previous work. }
\setcounter{equation}{0}

We will briefly recall our previous work \cite{bagi} on $\sigma$-models here. 

Let $\xi=(\xi_1,\xi_2)\in \C^2\backslash\{0\}$. Then
\be
S^3=\{z=\frac{\xi}{||\xi||}\;,\ \xi\in \C^2\backslash\{0\}\;,\ 
 ||\xi||=(\sum|\xi_\alpha|^2)^{1/2}\}\ .
\label{dsi}
\ee 
A point $\vec x\ (\vec x\cdot\vec x=1)$ on the sphere $S^2$ is
related to $z$ by
\be
x_i=z^\dagger\tau_iz\ ,\quad\tau_i=\ \hbox{Pauli 
matrices.} \label{dsii}
\ee
$S^2$ is the complex manifold $\CP^1$, with a
complex structure  inherited from that of $\C^2\backslash\{0\}$
with its complex variables $z$: the holomorphic coordinates on
$\CP^1$ are obtained from  $\C^2\backslash\{0\}$ by the projective maps
$\xi\to\frac{\xi_1}{\xi_2}=\frac{z_1}{ z_2}=\frac{x_1-ix_2}{1-x_3}$
(if $z_2\ne 0$, i.e. away from the north pole),
and $\to\frac{\xi_2}{\xi_1}=\frac{z_2}{ z_1}=\frac{x_1+ix_2}{1+x_3}$
(if $z_1\ne 0$,
i.e. away from the south pole).

A winding number $\kappa$ map to the target $S^2$ can be constructed 
as follows. Let $\kappa >0$ first, and the `partial isometry'
$v_\kappa$ be defined as
\be
v_\kappa:\ S^3\to S^3\ ,\quad v_\kappa(z)=
\frac{1}{\sqrt{ |z_1|^{2\kappa}+|z_2|^{2\kappa} }}
\left(\begin{matrix} z_1^{\kappa}\\ z_2^{\kappa}\\ \end{matrix}\right)\ .
\label{dsiii}
\ee
Note that as $z\ne 0$, not both $z_\alpha$ can be zero and hence
$\frac{1}{\sqrt{ |z_1|^{2\kappa}+|z_2|^{2\kappa} } }$ is well-defined.
 $v_\kappa(z)$ is of degree $\kappa$
 under $z\to e^{i\theta}z$, meaning that
 $v_\kappa(z)\to v_\kappa(z)e^{i\kappa\theta}$.

The field $n^{(\kappa)}$ on $S^2$ associated with $v_\kappa$ has components
 $n^{(\kappa)}_a$, that at  the point $\vec x$ take the values
\be
n^{(\kappa)}_a(\vec x)= v_\kappa(z)^\dagger\tau_a v_\kappa(z)\ .
\label{dsiv}
\ee
The invariance of the R.H.S. under $z\to z e^{i\theta}$ means that 
they depend just on $\vec x$.

 $v_{-|\kappa|}(\bar z)=\overline{ v_{|\kappa|}(z)}$ has degree
$-|\kappa|$. The construction of $n^{(\kappa)}$ for $\kappa=-|\kappa|$ 
is carried out using $\overline{ v_{|\kappa|}}$ for $v_{|\kappa|}$.

(\ref{dsiv})\ and its analogue for $\kappa <0$ give particular maps 
$S^2\to S^2$ with winding number $\kappa$. 
The general map ${\cal N}^{(\kappa)}$ for either
sign of $\kappa$ is got by replacing $v_\kappa(z),\ 
v_{\kappa}(z)^\dagger$ in (\ref{dsiv}) by
\be
{\cal V}^{(\kappa)}(z)=U(\vec x)v_\kappa(z)\ ,\quad 
{\cal V}^{(\kappa)\dagger}(z)=v_\kappa(z)^\dagger U(\vec x)^\dagger \ ,
\label{dsv}
\ee
where $U(\vec x)$ is any $2\times 2$ unitary matrix which depends only on 
$\vec x$.  In  our previous paper it was shown that the winding number of a map 
can be calculated by
\be
\kappa=\frac{1}{8\pi}\int_{S^2}\epsilon_{abc}{\cal N}^{(\kappa)}_a\,
d{\cal N}^{(\kappa)}_b\,d{\cal N}^{(\kappa)}_c=\frac{1}{2\pi i}\int_{S^2}
d({\cal V}^{(\kappa)\dagger}d{\cal V}^{(\kappa)})\ :
\ee
so long as $U$ is
a function of $\vec x$, the R.H.S. is equal to $\kappa$.

Fuzzy models of these maps are obtained from (\ref{dsiii}) by replacing 
$z_\alpha, \bar z_\beta$ by annihilation and creation operators
$a_\alpha,\; a^\dagger_\beta :\ [a_\alpha, a^\dagger_\beta]=
\delta_{\alpha\beta}$ etc..
Then $v_\kappa\to\hat v_\kappa$, where for $\kappa>0$
\beqa
\hat v_\kappa&=&\left(\begin{matrix}a^\kappa_1\\ a^\kappa_2\\ \end{matrix}
\right)
\frac{1}{\sqrt{\hat Z_\kappa}}\ ,\quad
\hat v^\dagger_\kappa=\frac{1}{\sqrt{\hat Z_\kappa}}
\left(\begin{matrix}(a^\dagger_1)^\kappa\ , &(a^\dagger_2)^\kappa\\ 
\end{matrix}\right)\ ,  \label{dsvi}             \\
\hat Z_\kappa&=&\hat Z_\kappa^{(1)}+\hat Z_\kappa^{(2)}\ ,\quad
\hat Z_\kappa^{(\alpha)}=\hat N_\alpha(\hat N_\alpha-1)...
(\hat N_\alpha-\kappa+1)
\ ,\quad\hat N_\alpha=a^\dagger_\alpha a_\alpha  \ ,\nn
\eeqa
while for $\kappa<0$, we change $\hat v_\kappa$ to
\beqa
\hat v_{-|\kappa|}&=&\left(\begin{matrix}(a_1^\dagger)^{|\kappa|}\\
(a_2^\dagger)^{|\kappa|}\\ \end{matrix}\right)
\frac{1}{\sqrt{\hat Z_{-|\kappa|}}}\ ,\quad
\hat v_{-|\kappa|}^\dagger=
\frac{1}{\sqrt{\hat Z_{-|\kappa|}}}\left(\begin{matrix}a_1^{|\kappa|}\;,&
a_2^{|\kappa|}\\ \end{matrix}\right)    \label{dsvii}         \\
\hat Z_{-|\kappa|}&=&\hat Z_{-|\kappa|}^{(1)}+\hat Z_{-|\kappa|}^{(2)}\ ,\quad
\hat Z_{-|\kappa|}^{(\alpha)}=(\hat N_\alpha+|\kappa|)
(\hat N_\alpha+|\kappa|-1)...
(\hat N_\alpha+1)
\ ,\quad\hat N_\alpha=a^\dagger_\alpha a_\alpha  \ .\nn
\eeqa
The particular expressions chosen  for $\hat Z_\kappa$  ensure
the normalization $\hat v_\kappa^\dagger \hat v_\kappa=1$.

With this definition we can use (\ref{dsvi}, \ref{dsvii}) for either
sign of $\kappa$. 

The quantized $\hat x$ and $\hat n^{(\kappa)}$ for either sign of $\kappa$ are
\be
\hat x=\hat v^\dagger_1\,\vec\tau\,\hat v_1\quad,\quad \hat n^{(\kappa)}=
\hat v^\dagger_\kappa\,\vec\tau\,\hat v_\kappa \ .
\label{dsviii}
\ee

The general fuzzy $\sigma$-field $\hat{\cal N}^{(\kappa)}$ is obtained
by transforming $\hat v_\kappa$ by a $2\times 2$ unitary matrix $\hat U$,
where $\hat U_{\alpha\beta}$ is a function of $a^\dagger_\alpha a_\beta$.
That gives 
\be
\hat{\cal V}^{(\kappa)}=\hat U\hat v_\kappa\quad,\quad
\hat{\cal N}^{(\kappa)}_a=\hat{\cal V}^{(\kappa)\dagger}\tau_a
\hat{\cal V}^{(\kappa)}
\label{dsix}
\ee

Now $\hat x,\;\hat{\cal N}^{(\kappa)}$ commute with the number operator
$\hat N=\hat N_1+\hat N_2$ and hence can be restricted to the subspace
$\hat N=n\;(>0)$. This subspace has dimension $n+1$. That gives us a finite
dimensional matrix model for fuzzy solitons.

%
\section{ A Generalization. }
\setcounter{equation}{0}

To motivate a generalization of (\ref{dsiii}), let us note that
taking as complex 
coordinates $\zeta=\frac{z_2}{z_1}=\frac{x_1+ix_2}{1+x_3}$ 
in a patch of $S^2$, and
$\phi=\frac{n_1+in_2}{1+n_3}=\frac{v_2}{v_1}$  in a patch of target $S^2$,
 one can express the 'energy density' of a field configuration by
\be
-{\cal L}_in_a{\cal
  L}_in_a=-\frac{4{\cal L}_i\phi{\cal L}_i\bar\phi}{(1+\phi\bar\phi)^2}=
2\frac{(1+\zeta\bar\zeta)^2}{(1+\phi\bar\phi)^2}\Big(\frac{\partial\phi}{\partial\zeta}
\frac{\partial\bar\phi}{\partial\bar\zeta}+
\frac{\partial\phi}{\partial\bar\zeta}\frac{\partial\bar\phi}{\partial\zeta}\Big)
\ ,\quad {\cal L}_i=-i\epsilon_{ijk}x_j\partial_k\ .
\label{dsxi}
\ee
  The 'energy density' of the field configuration (\ref{dsiv}), which has 
$\phi(\zeta)=\zeta^\kappa$, is
  therefore concentrated around the north pole $x_3=1$ of the sphere.
For a soliton of variable width and height
localized at
\be 
\zeta'=\frac{x'_1+ix'_2 }{ 1+x'_3}\ ,
\label{dsxii}
\ee
we can change (\ref{dsiii}) to 
\be
v_\kappa(z,\zeta',\lambda)=
\frac{1}{\sqrt{|\lambda z_1|^{2\kappa}+|z_2-\zeta' z_1|^{2\kappa}}}
\left(\begin{matrix}(\lambda z_1)^{\kappa}\\
(z_2-\zeta' z_1)^{\kappa}\\ \end{matrix}
\right)\quad,\ \lambda\ne 0
\label{dsxiii}
\ee
for $\kappa>0$. 
More generally, a multisoliton field configuration with winding number
$\kappa$ can be obtained replacing the partial isometry (\ref{dsiii}) with 
one of the form
\be 
 v_\kappa(z,c)=\frac{1}{\sqrt{|P_{1\kappa }|^2+|P_{2\kappa }|^2}}
\left(\begin{matrix}P_{1\kappa }(z)\\ P_{2\kappa }(z)\\
\end{matrix}\right)\ , \quad P_{ \alpha \kappa}(z)=
\sum_{h=0}^\kappa c_{\alpha h}z_1^{\kappa-h}z_2^h\ ,
\label{dsxiiib}
\ee
with arbitrary complex coefficients $c_{\alpha h},\;\alpha=1,2,\ h=0,...,\kappa$,
such that $P_{1\kappa}(z),\;P_{2\kappa}(z)$ have no common
zeroes on $S^2$ (so that the denominator  is not $0$ on $S^2$). This partial 
isometry can be obtained from the $v_\kappa(z)$ defined in (\ref{dsiii}) applying to it $V(\vec x)= 
v_\kappa(z,c)\,v_\kappa(z)^\dagger$, which is a function of $\vec x$ as 
indicated. Therefore, if we use  $v_\kappa(z,c)$ and then a unitary
$U(\vec x)$ to construct 
$n^{(\kappa)}(\vec x,c),\; {\cal V}_\kappa(z,c)$ 
and ${\cal N}^{(\kappa)}(\vec x,c)$, we will find that
\begin{itemize}
\item $n^{(\kappa)}(\vec x,c)$ is still invariant under
$z\to e^{i\theta}z$, and hence depends only on $\vec x$.

\item By the argument indicated after (\ref{dsv}), the winding number of 
$n^{(\kappa)}(\vec x,c)$ is indeed $\kappa$.

\item  By the same argument, as long as $U$ is a function of $\vec x$,  
 ${\cal N}^{(\kappa)}(\vec x,c)$ depends just on $\vec x$ and has
 winding number  $\kappa$.
\end{itemize}

If $c_{\alpha\,\kappa}\ne 0$ for example, the 
coordinate $\phi$ introduced above for a patch of target $S^2$ can be
written for this type of configurations in the form
\be
\phi(\zeta,c)=
c\frac{(\zeta-a_1)...(\zeta-a_\kappa)}{(\zeta-b_1)...(\zeta-b_\kappa)}\ ,
\quad c=\frac{c_{2\kappa}}{c_{1\kappa}}\ne 0\ ,
\label{dsxiiic}
\ee
which is the one given in \cite{belavin}, \cite{fateev}.

For $\kappa<0$ we let $v_{-|\kappa|}(z,c)=
 v_{|\kappa|}(\bar z,c)$, and then
construct the rest in an obvious manner.

For quantization, we let $z_\alpha\to a_\alpha,\;
\bar z_\alpha\to a^\dagger_\alpha$, but keep $c_{\alpha h}$ as 
complex numbers. Then for example for $\kappa>0$,
\be
\hat v_\kappa(c)=\hat w^{(\kappa)}(c)
\frac{1}{ (\hat w^{(\kappa)\dagger}(c)\hat w^{(\kappa)}(c))^{1/2}}
\quad,\quad \hat w^{(\kappa)}(c)=
\left(\begin{matrix}
\sum c_{1h}a_1^{\kappa-h}a_2^h\\ \sum c_{2h}a_1^{\kappa-h}a_2^h
\\ \end{matrix}\right)\ .
\label{dsxiv}
\ee
In this way we get all the fuzzy solutions with 
$c_{\alpha h}$-dependence. 

Note that we can study fuzzy solitons using
the expressions in (\ref{dsxiv}), which are well-defined. There is no
need to find analogues of (\ref{dsxiiic}), which at best would be messy.

%
\section{ Duality for commutative $\  {\rm CP}^1\to {\rm CP}^1 $.   }
\setcounter{equation}{0}
 In the $\CP^1$-model on $S^2$, without any discretization, self--duality and
anti--self--duality are the conditions
\be
{\cal L}_i{\cal N}_a^{(\kappa)}\mp \epsilon_{ijk}\,x_k\;\epsilon_{abc}\,
{\cal N}_c^{(\kappa)}\,{\cal L}_j{\cal N}_b^{(\kappa)}=0 \ .
\label{dsxxi}
\ee
It requires some work to show that
\begin{itemize}
\item Self--dual solutions require $\kappa>0$ and are given by the choice
$U(\vec x)=1,\ {\cal N}_a^{(\kappa)}(\vec x,c)=
n_a^{(\kappa)}(\vec x,c)$.
\item Anti--self--dual solutions require $\kappa<0$ and are given by 
the choice $U(\vec x)=1,\ {\cal N}_a^{(\kappa)}(\vec x,c) =
n_a^{(\kappa)}(\vec x,c)$.
\end{itemize}

 Belavin and Polyakov recognized that the self-duality or anti-self-duality
conditions are equivalent to Cauchy-Riemann equations on the world
sheet $\R^2$. Summarizing their argument, one may use the complex coordinate 
$\phi$ for the target $S^2$ introduced in the
previous section to find that (\ref{dsxxi}) implies
\be
{\cal L}_i\phi=\pm i\,\epsilon_{ijk}x_j\,{\cal L}_k\phi\ ,
\label{dsxxia}
\ee
and that, in terms of the complex coordinate $\zeta$, this is equivalent to
\be
\frac{\partial\phi}{\partial\bar\zeta}=0\quad \hbox{for the upper sign},
\quad\quad 
\frac{\partial\phi}{\partial\zeta}=0\quad \hbox{for the lower sign}.
\label{dsxxic}
\ee

To translate  the conditions to the fuzzy case we need however a
statement which does not involve ratios or local coordinates.
This we provide in \S 5. 

%
\section{ Analyticity and  Duality. }
\setcounter{equation}{0}

The analyticity properties of duality equations can be partly attributed to a certain scale
invariance of the latter. We shall first introduce a formalism which
explicitly brings out this invariance.

Let 
\be
D^cSO(3)=\{s\in Mat_2(\C):\ s^\dagger s=\Delta\BI,\ \Delta>0\}\ .
\label{dsxxxi}
\ee
It is clear that if $\Delta\ne 0$, then $\Delta>0$, and that $D^cSO(3)$ is a
group. It is the central extension of $SO(3)$ by complex dilatations
$D^c$.
We can show this as follows. First we quotient $D^cSO(3)$ by the
connected component of real dilatations
\be
D_0^R=\{\lambda\BI:\ \lambda>0\}\ ,
\label{dsxxxia}
\ee
to get $U(2)$, the homomorphism $D^cSO(3)\to U(2)$ being
\be
s\ \to\ s\,\frac{1}{ |s|}\quad,\quad |s|=(s^\dagger s)^{1/2}\,>\,0\ .
\label{dsxxxii}
\ee
Its kernel consists of positive multiples of $\BI$, that is
of $D_0^R$. But  $D^cSO(3)$ contains also $U(1)=\{\alpha\BI:\
|\alpha|=1\}$.
On quotienting $U(2)$ by $U(1)$ we get $SO(3)$.

A map from $\C^2\backslash\{0\}$ to $D^cSO(3)$, and then to $S^2$ is given by
\beqa
\C^2\backslash \{0\}\ \ni\ \xi\ &\to&\  
s=\left(\begin{matrix} \xi_1&-\bar\xi_2\\\xi_2&\bar\xi_1\\ \end{matrix}
\right)\in D^cSO(3)\ ,\nonumber \\
s\ &\to&\ u(s)= s\,\frac{1}{|s|}\in U(2)\ ,\nonumber\\
u(s)\ &\to&\ u(s)\,\tau_3\,u(s)^{-1}= \tau_i\,(Ad\,
u(s))_{i3}:=\tau_ix_i\in S^2\ ,
\label{alti}
\eeqa
where $Ad\,u(s)$ is the matrix of $u(s)$ in the adjoint
representation. Note that since $s\tau_3 s^{-1}=u(s)\tau_3 u(s)^{-1}$,
$Ad\,s$ is equal to $Ad\,u(s)$.

Under infinitesimal rotations $\xi\to\xi'=(\BI+\frac{i}{2}\alpha_i\tau_i)\xi$,
$s\to s'=(\BI+\frac{i}{2}\alpha_i\tau_i)s$, $x_i\to x'_i=x_i-
\alpha_k\epsilon_{kij}x_j$.  Functions on $S^2$ can be pulled back to 
functions on  $D^cSO(3)$, and by comparing the actions of rotations, we find that the 
angular momentum generators ${\cal L}_i$ can be lifted to the negative of
the left acting $L_i$'s, defined by 
\be
(e^{i\alpha_jL_j}f)(s)=f(e^{-i\alpha_j\tau_j/2}\,s)\ .
\label{dsxlviii}
\ee

Consider maps ${\cal W}_{\kappa}:\ 
\C^2\backslash\{0\}\to \C^2\backslash\{0\}$ which give
$ {\cal V}_\kappa$ on normalization. For $\kappa>0$ we have taken 
\be
{\cal W}_\kappa(\xi)= U(\vec x)w_\kappa(\xi)
\quad,\quad w_\kappa(\xi)=\left(\begin{matrix}\xi_1^\kappa\\
\xi_2^\kappa\\ \end{matrix}\right)\ ,
\label{dsxxii}
\ee
while for $\kappa<0$, we must complex conjugate the $\xi_\alpha$-s in
 $w_\kappa(\xi)$.
Now ${\cal W}_\kappa$ gives us a set of maps 
${\cal G}=\{g\}$ from $\C^2\backslash\{0\}$
to $D^cSO(3)$:
\be
g(\xi)=
\left(\begin{matrix}{\cal W}_{\kappa\,1}&-{\overline{\cal W}}_{\kappa\,2}\\
  {\cal W}_{\kappa\,2}& {\overline{\cal W}}_{\kappa\,1}\\ \end{matrix}\right)
(\xi)\quad ,\quad
g^\dagger g={\cal W}_\kappa^\dagger{\cal W}_\kappa\,\BI
\ .\label{dsxxxiv}
\ee
$g(\xi)$ is indeed valued in $D^cSO(3)$ since ${\cal W}_{\kappa\, 1,2}$
have no common zeroes in $\xi$ and hence ${\cal W}_\kappa^\dagger{\cal
  W}_\kappa>0$ for all $\xi$. These maps are such that
\be
g\,\tau_3\,g^{-1}=\tau_a{\cal N}^{(\kappa)}_a
\label{altii}
\ee
Associating an ${\cal R}\in SO(3)$ to $g$ in the usual way by
\be
g\,\tau_a\,g^{-1} =
\tau_b\,{\cal R}_{ba}\ ,
\label{dsxliii}
\ee
we see that ${\cal N}^{(\kappa)}_a={\cal R}_{a3}$. 

Eq. (\ref{dsxliii}) is invariant under scale transformations of $g$. Now
(\ref{dsxxi}) can be expressed in terms of $g$, as we shall show below. Thus 
it can depend only on ratios of components of $g$, an important result
 (cf. (\ref{dsxiiic})). The formalism using $g$ is convenient to express 
this scale invariance.

Indicating by $\theta_a,\ (\theta_a)_{ij}=-i\epsilon_{aij}$, the components
of spin 1 angular momentum, we have from (\ref{alti}), (\ref{altii}) that:
\be
\big((Ad\,s)\theta_3(Ad\,s)^{-1}\big)_{ij}=-i\epsilon_{ijk}x_k\quad,\quad
\big({\cal R}\theta_3{\cal R}^{-1}\big)_{ab}=-i\epsilon_{abc}
{\cal N}^{(\kappa)}_c
\label{altiii}
\ee
We may therefore rewrite (\ref{dsxxi}) in the form
\be
(Ad\, s)^{-1}_{ij}({\cal R})^{-1}_{ab}L_j{\cal R}_{b3}=
\mp\big(\theta_3(Ad\, s)^{-1}\big)_{ij}\big(\theta_3{\cal R}^{-1}\big)_{ab}
L_j{\cal R}_{b3}\ .
\label{altiv}
\ee
Next we can go from the left-acting $L_i$ to the right acting $L^R_i$,
defined by
\be
(e^{i\alpha_jL_j^R}f)(s)=f(s\,e^{i\alpha_j\tau_j/2})\ ,
\label{altv}
\ee
using the relation
\be
(Ad\, s)^{-1}_{ij}L_j=-L_i^R\ ,
\label{altvi}
\ee
thus turning (\ref{altiv}) to
\be
({\cal R}^{-1}\,L_i^R\,{\cal R})_{a3}=\mp(\theta_3)_{ij}
(\theta_3\,{\cal R}^{-1}\,L_j^R\,{\cal R})_{a3}\ .
\label{altvii}
\ee
Taking successively $i=3,1,2$, we see that the independent relations we
have are
\be 
({\cal R}^{-1}\,L_3^R\,{\cal R})_{a3}=
({\cal R}^{-1}\,L_-^R\,{\cal R})_{13}\mp i
({\cal R}^{-1}\,L_-^R\,{\cal R})_{23}=0\ ,
\label{altviii}
\ee
where $L_-^R=L_1^R- iL_2^R$. The real and imaginary parts of (\ref{altviii}) give 
(\ref{altvii}), because  ${\cal R}_{ab}$ is real, 
and $(L^R_j{\cal R})_{ab}$
is pure imaginary. The matrices $({\cal R}^{-1}\,L_i^R\,{\cal R})$ are
therefore antisymmetric, and since the $\theta_a$ are pure imaginary,
these equations are equivalent to:
\be
\tr\big((\theta_1-i\theta_2)\,{\cal R}^{-1}\,L_3^R\,{\cal R}\big)=
\tr\big((\theta_1\mp i\theta_2)({\cal R}^{-1}\,L_-^R\,{\cal
  R})\big)=0\ .
\label{altix}
\ee
These relations must  hold in any representation of the Lie algebra, and
therefore
\be
\tr\big((\tau_1-i\tau_2)\,g^{-1}L^R_3g\big)=
\tr\big((\tau_1\mp i\tau_2)\, g^{-1}L^R_-g\big)=0\ .
\label{altx}
\ee
In this way we have succeeded in expressing the self-duality - 
anti-self-duality conditions (\ref{dsxxi}) directly in terms of the 
${\cal W}_{\kappa\,\alpha},\;\overline{\cal W}_{\kappa\,\alpha}$. Explicitly, the
relations we have obtained are:
\beqa
\overline{\cal W}_{\kappa\,1}L_3^R\overline{\cal W}_{\kappa\,2}-
\overline{\cal W}_{\kappa\,2}L_3^R\overline{\cal W}_{\kappa\,1}&=&0
\label{altxi}\\
\hbox{upper sign:}\ -\overline{\cal W}_{\kappa\,1}L_-^R\overline{\cal W}_{\kappa\,2}
+\overline{\cal W}_{\kappa\,2}L_-^R\overline{\cal W}_{\kappa\,1}&=&0 \nn\\\
\hbox{lower sign:}\ \ 
-{\cal W}_{\kappa\,2}L_-^R{\cal W}_{\kappa\,1}+{\cal W}_{\kappa\,1}L_-^R{\cal W}_{\kappa\,2}&=&0
\label{altxii}
\eeqa
First consider (\ref{altxii}), and for definiteness the 'upper sign'
(self-duality) equation. What it means is that
the ratio $\frac{\overline{\cal W}_{\kappa 2}}
{\overline{\cal W}_{\kappa 1}}$
 is annihilated by $L_-^R$.
Now on functions $f$ on $D^cSO(3)$,
\be
(L_-^Rf(s)=(\bar\xi_1\frac{\partial}{\partial\xi_2}
-\bar\xi_2\frac{\partial}{\partial\xi_1})f(s)\ .
\label{altxiiia}
\ee
Hence  $\frac{\overline{\cal W}_{\kappa 2}}
{\overline{\cal  W}_{\kappa 1}}$ depends only on $(\bar\xi_1,\bar\xi_2)$. 
This implies that the dependance of 
$({\cal W}_{\kappa\,1},{\cal W}_{\kappa\,2})$ on $(\bar\xi_1,\bar\xi_2)$ factors out when we
take the ratio, or can be eliminated by a rescaling.
Suppose we do this rescaling, so that we may represent the ${\cal W}_{\kappa\,\alpha}$'s
in the form ${\cal W}_{\kappa\,\alpha}=\sum_{kn} c_{\kappa\alpha k  n}\xi^{k-n}_1\xi_2^n$.
Then, since we may express $L_3^R$ as
\be
(L_3^Rf)(s)=\half\sum_{\alpha=1,2}
\big(\xi_\alpha\frac{\partial }{\partial \xi_\alpha}-
\bar\xi_\alpha\frac{\partial }{\partial \bar\xi_\alpha}\big)f(s)\  ,
\label{altxiii}
\ee
 (\ref{altxi})  implies that the sum must be restricted to a single value
of $k$, the same for both values of $\alpha$. It follows that for
 self-duality we must have
%
%
%
%
%
\be
{\cal W}_{\kappa\,1}=\sum_{n=0}^\kappa c_{1 n}\xi^{\kappa-n}_1\xi_2^n\quad,\quad
{\cal W}_{\kappa\,2}=\sum_{n=0}^\kappa c_{2 n}\xi^{\kappa-n}_1\xi_2^n \ .
\label{altxxix}
\ee
for some integer $\kappa>0$ and coefficients $c_{\alpha n}$.

We can interpret (\ref{altxxix}) by saying that ${\cal
  W}_{\kappa\,\alpha}$ are highest weight vectors with angular
  momentum $\kappa/2$ for the $SU(2)$ Lie algebra generated by
  $L^R_i$. They are holomorphic (being polynomials) in $\xi_\alpha$.

A similar discussion can be made for the lower sign in (\ref{altxii})
(anti-self-duality).

Summarizing, (\ref{dsxxi}) expresses duality using unit vectors of world sheet
and target $S^2$. That is not the best way for fuzzy physics. 
For the latter, it
is better to rewrite it after scaling and using the holomorphic
coordinates of $\C^2\backslash\{0\}$.

%
 \section{ Fuzzification of  Duality. }
\setcounter{equation}{0}

The dual solutions for $\CP^1\to\CP^1$ are $v_\kappa(z,c)$ 
and their derived structures. They are very easy to quantize: replace 
$\xi_\alpha$ by $a_\alpha$ and $\bar\xi_\alpha$ by $a_\alpha^\dagger$. 
That gives a fuzzy version of BPS states.

In the commutative case, BPS solutions saturate the lower bound on the energy
functional \cite{belavin}. Such a result is not quite correct in the fuzzy
case \cite{bagi}.

\begin{acknowledgments}

We are deeply indebted to T. R. Govindarajan and E. Harikumar for
drawing our attention to the formulation of BPS solitons in fuzzy
physics, explaining their approach to fuzzy sigma models and many
discussions. 
This work was supported by DOE  under contract numbers 
DE-FG02-85ER4023, by NSF  under contract numbers INT-9908763, and
by INFN and MIUR, Italy.
\end{acknowledgments}

\end{document}